%% file: main_file.tex
\documentclass[conference]{IEEEtran}
\usepackage{cite}
\usepackage{amsmath,amssymb,amsfonts}
\usepackage{algorithmic}
\usepackage{graphicx}
\usepackage{textcomp}
\usepackage{listings}
\usepackage{amsmath}
\usepackage{multirow}
\usepackage{enumitem}
\usepackage{hyperref}

\def\BibTeX{{\rm B\kern-.05em{\sc i\kern-.025em b}\kern-.08em
    T\kern-.1667em\lower.7ex\hbox{E}\kern-.125emX}}

\usepackage{ifthen}
\usepackage{balance}
\newboolean{showcomments}
\setboolean{showcomments}{true}
\ifthenelse{\boolean{showcomments}}
{ \newcommand{\mynote}[2]{\textcolor{red}{
    \fbox{\bfseries\sffamily\scriptsize#1}
    {\small$\blacktriangleright$\textsf{\emph{#2}}$\blacktriangleleft$}}}}
{ \newcommand{\mynote}[2]{}}

\usepackage[dvipsnames]{xcolor}
  \definecolor{diffstart}{named}{Blue}%{Grey}
  \definecolor{diffincl}{named}{OliveGreen}
  \definecolor{diffrem}{named}{Red}

\newcommand{\extrabold}{\bfseries}

\usepackage{listings}
  \lstdefinelanguage{diff}{
	basicstyle=\ttfamily\extrabold\tiny,
	morecomment=[f][\color{diffstart}]{@},
	morecomment=[f][\color{diffincl}]{+},
	morecomment=[f][\color{diffrem}]{-},
        numbers=left,
        stepnumber=1,
        keepspaces=true,
	identifierstyle=\color{black},
  }

\begin{document}

% \title{Hierarchical Learning of Patch Embedding for Stable Patch Identification in Linux Kernel}
% \title{PatchNet: Hierarchical Deep Learning-Based Stable Patch Identification for the Linux Kernel}
\title{PatchNet: A Tool  for Deep Patch Classification}
% {\footnotesize \textsuperscript{*}Note: Sub-titles are not captured in Xplore and
% should not be used}
% \thanks{Identify applicable funding agency here. If none, delete this.}
% }
%
\author{\IEEEauthorblockN{Thong Hoang\textsuperscript{1}, Julia Lawall\textsuperscript{2}, Richard J. Oentaryo\textsuperscript{4}, Yuan Tian\textsuperscript{3}, David Lo\textsuperscript{1}}
\IEEEauthorblockA{\textsuperscript{1}Singapore Management University/Singapore, \textsuperscript{2}Sorbonne University/Inria/LIP6\\
\textsuperscript{3}Queen's University/Canada, \textsuperscript{4}McLaren Applied Technologies/Singapore}
vdthoang.2016@smu.edu.sg, Julia.Lawall@lip6.fr, richard.oentaryo@mclaren.com, \\ yuan.tian@cs.queensu.ca, davidlo@smu.edu.sg
}

% \author{\IEEEauthorblockN{Thong Hoang\textsuperscript{1}, Julia Lawall\textsuperscript{2}, David Lo\textsuperscript{1}}
% \IEEEauthorblockA{\textsuperscript{1}School of Information Systems, Singapore Management University, Singapore\\
% \textsuperscript{2}Inria/LIP6, Regal}
% vdthoang.2016@smu.edu.sg, julia.lawall@lip6.fr, davidlo@smu.edu.sg
% }

\maketitle

\input{abstract}

%\begin{IEEEkeywords}
%Linux kernel, bug fixing, classification, deep learning.
%\end{IEEEkeywords}

\input{intro}

\input{design}
\input{preproc}

\input{usage}
\input{conclusion}

\input{ack}

\balance
\bibliographystyle{IEEEtran}
\bibliography{references} 
\end{document}

%% file: abstract.tex
\begin{abstract}
This work proposes PatchNet, an automated tool based on hierarchical deep
learning for classifying patches by extracting features from commit messages
and code changes. PatchNet contains a deep hierarchical structure that
mirrors the hierarchical and sequential structure of a code change,
differentiating it from the existing deep learning models on source
code. PatchNet provides several options allowing users to select parameters
for the training process. The tool has been validated in the context of
automatic identification of stable-relevant patches in the Linux kernel and
is potentially applicable to automate other software engineering tasks that
can be formulated as patch classification problems. A video demonstrating PatchNet is available at~\url{https://goo.gl/CZjG6X}. The PatchNet
implementation is available at~\url{https://github.com/hvdthong/PatchNetTool}.
% Our video demonstrating PatchNet
% %demonstration on the performance of PatchNet 
% and PatchNet implementation are publicly available at~\url{https://goo.gl/CZjG6X} and~\url{https://github.com/hvdthong/PatchNetTool} respectively.
\end{abstract}

%% file: intro.tex
\section{Introduction and Related Work}
\label{sec:intro}

Deep learning techniques have recently been used to automate some software
engineering tasks such as code clone
detection~\cite{white2016deep}, software traceability link
recovery~\cite{guo2017semantically}, and bug
localization~\cite{lam2017bug}. However, no existing work
has investigated the problem of learning a semantic representation of code
changes, {\em i.e.},
patches, for classifying them into predefined classes. Patches are composed
of a short text describing a change, 
% the code before the change, the change, and the code after the change, 
the lines removed by the change, and the lines added by the change, and all these pieces need to be considered in a
holistic way to produce a good semantic representation. Patch
classification is an important problem since many automated software
engineering tasks, such as just-in-time defect
prediction~\cite{kamei2016studying}, tangled change
prediction~\cite{kirinuki2014hey}, code reviewer recommendation for a
commit~\cite{rahman2016correct}, etc. can be mapped to it.

Close to our work on patch classification is the prior work by Tian et
al.~\cite{tian2012identifying} that proposes an automated patch
classification approach integrating 
% two different algorithms 
LPU (Learning
from Positive and Unlabeled Examples)~\cite{letouzey2000learning} and SVM
(Support Vector Machine) to build a classification
model. To apply LPU+SVM to patches, Tian et al. manually defined
a set of features extracted from patches. However, this manual creation
process may overlook features that are helpful to classify patches. The
chosen features are also specific to a particular patch classification
setting (i.e., bug fixing patch identification), and a new set of features
is likely needed for other settings. In this work, we replace this manual
feature engineering step by leveraging the power of deep learning. In
particular, we construct a model that can extract a good semantic
representation capturing a patch's hierarchical and structural properties.

This paper presents our tool PatchNet that
performs learning on a set of patches. PatchNet performs deep patch
classification in two phases. In the training phase, it takes as input a
set of labeled patches to learn a deep learning model. This model is then
used in the prediction phase on a set of unlabeled patches to produce
scores that estimate how likely the given patches fit the class
labels. Specifically, PatchNet aims to automatically learn two embedding
vectors, representing a commit message and a set of code changes in a given
patch, respectively. The two embedding vectors are then used to compute a
prediction score from a given patch estimating how likely the patch is
relevant to a particular class.

PatchNet is implemented as a command line tool. In the training phase, the
user provides a file of labeled patches and the name of a folder
in which to put the trained PatchNet model. In the prediction phase, the
command used to produce the prediction scores also takes two inputs:
a file of unlabeled patches and the name of the folder containing
the previously trained model. PatchNet also provides several supplementary options to
allow users to select hyperparameters for the training process. PatchNet
targets binary classification tasks, however, it can also be used for multi-label classification tasks by reducing the problem of multi-label classification to multiple binary classification problems~\cite{Bishop:2006:PRM:1162264}. PatchNet
currently only supports patches on C code.

PatchNet has been applied to the task of stable patch identification in the
Linux kernel. Specifically, the Linux kernel follows a two-tiered release
model in which a \textit{mainline} version accepts bug fixes and feature
enhancements, and a series of stable versions accepts only bug
fixes~\cite{lee2003firm}. While the mainline targets users who want to
benefit from the latest features, the stable versions target users who
value the stability of their kernel. The stable kernel
development process requires that all patches be submitted to the mainline,
and then propagated by maintainers from there to review for inclusion in
stable kernels. The wide variation in rates at which patches are
propagated by maintainers to stable kernels across the code base suggest
that some bug-fixing patches may be being overlooked. There is thus the
potential for an automated, learning-based approach to improve this
process. Our evaluation on 82,403 recent Linux patches shows that PatchNet
outperforms the state-of-the-art baseline (i.e., LPU+SVM), achieving
precision of 0.839 and recall of 0.907.  By releasing our
tool, we hope to enable others to apply PatchNet to other patch
classification tasks.

% The rest of this paper is organized as follows. Section~\ref{sec:design}
% provides a bird's-eye view of the PatchNet
% architecture. Section~\ref{sec:data} describes how PatchNet preprocesses
% the patch data before initiating the learning
% process. Section~\ref{sec:usage} illustrates how PatchNet can be used in
% various scenarios. Section~\ref{sec:exp} summarizes our previous experiment
% with PatchNet, that is described in more detail in the ICSE 2019
% technical paper track submission.  Finally, we conclude in Section~\ref{sec:conclusion}.

The rest of this paper is organized as follows. Section~\ref{sec:design}
provides a bird's-eye view of the PatchNet
architecture. Section~\ref{sec:data} describes how PatchNet preprocesses
the patch data before initiating the learning
process. Section~\ref{sec:usage} illustrates how PatchNet can be used in
various scenarios. 
% Section~\ref{sec:exp} summarizes the PatchNet's experiments. 
Section~\ref{sec:exp} summarizes some experiments with PatchNet.
Finally, we conclude in Section~\ref{sec:conclusion}.

%% file: design.tex
\section{Overall Design}
\label{sec:design}
\begin{figure*}[t!]
	\center
	\includegraphics[scale=0.33]{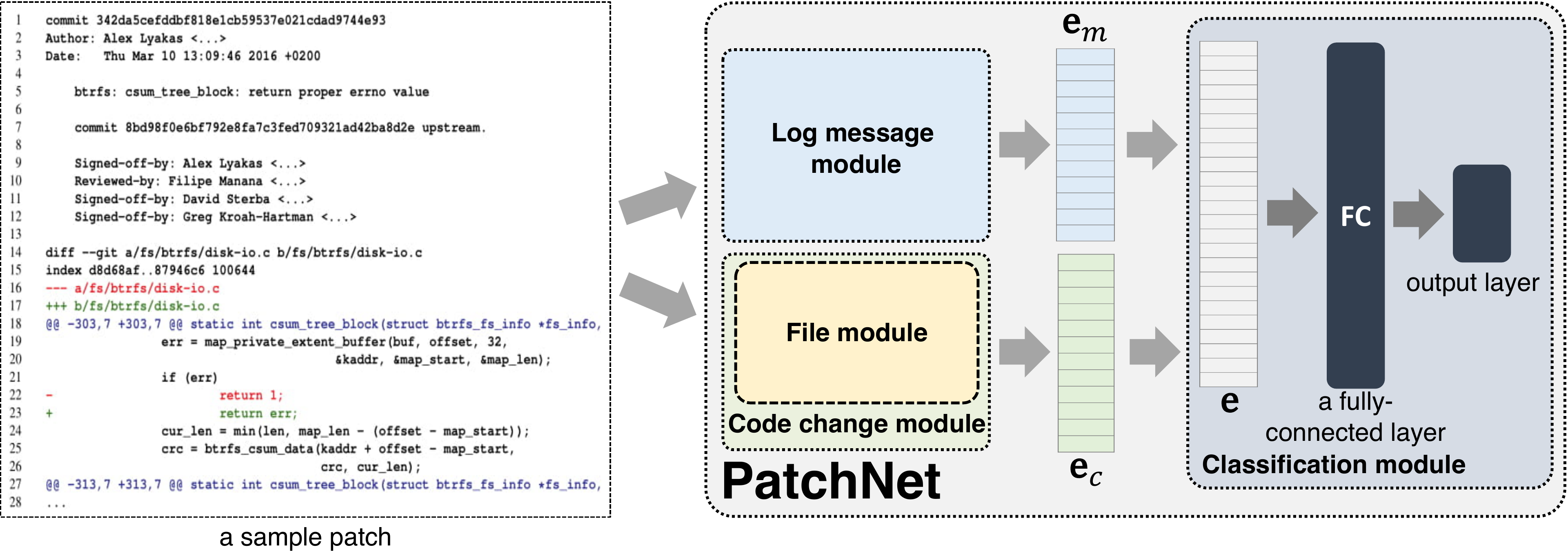}
	\vspace{-0.1cm}
	\caption{The overall design of PatchNet to classify stable vs. non-stable patches. A sample patch contains both a textual commit message (lines 5-12) and a set of code changes (lines 14-28) that are applied to an affected file. $\textbf{e}_m$ and $\textbf{e}_c$ are embedding vectors collected by the commit message module and code change module, respectively. $\textbf{e}$ is a single vector formed by concatenating these two embedding vectors.}
	\label{fig:patchnet}
    
\end{figure*}

\begin{figure}[t!]
	\center
	\includegraphics[scale=0.40]{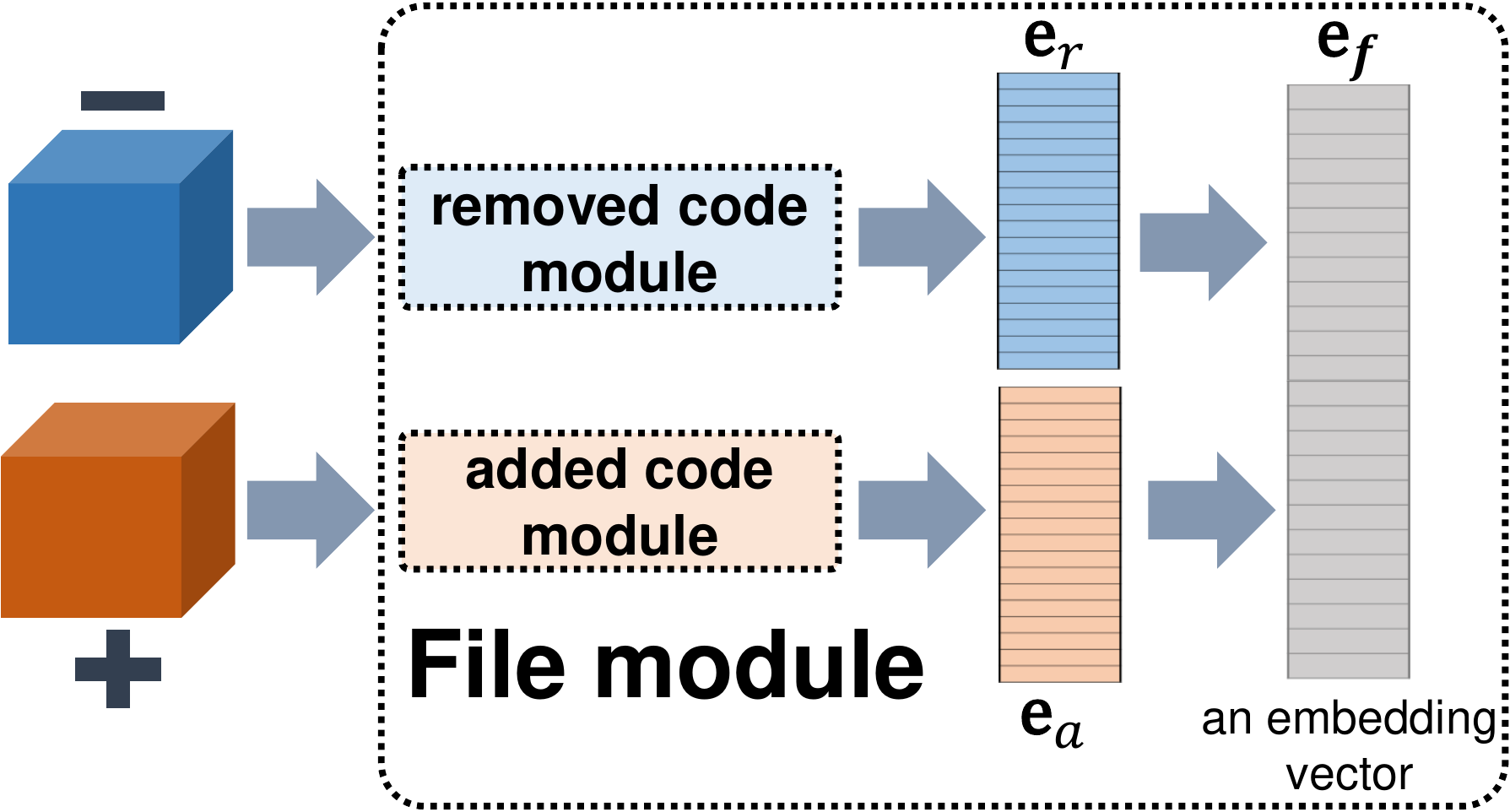}
	\vspace{-0.1cm}
	\caption{Architecture of the \textit{file module} for
          mapping a file in a given patch to an embedding vector. The input
          of the module is the removed code and added code of the affected
          file, denoted by ``--'' and ``+'', respectively.}
	\label{fig:commit_code_model}
\end{figure}

Figure~\ref{fig:patchnet} shows the overall design of PatchNet. PatchNet
accepts as input a set of patches, each of which contains both a commit message and a code change. The output of PatchNet associates each patch with
a prediction score reflecting how likely the patch satisfies the
criteria of the classification.
Our framework includes three main
modules: the \textit{commit message module}, the \textit{code change module}, and the \textit{classification module}. The commit message
module and the code change module transform the commit message and the
code change into embedding vectors $\text{e}_m$ and $\text{e}_c$,
respectively. The two embedding vectors are then passed to the
classification module, which computes the prediction score.
In the rest of this section, we give an overview of each of these modules.

\noindent \textbf{Commit message module:} The architecture of the commit message module is the same as the one proposed by Kim~\cite{kim2014convolutional} for sentence classification. This module takes a commit message as input and outputs an embedding vector that represents the most salient features of the message. Specifically, we encode the commit message as a two-dimensional matrix by viewing the message as a sequence of vectors where each vector represents a word appearing in the message. We then apply a convolutional layer followed by a max-pooling operation to obtain the message's salient features. 

\noindent \textbf{Code change module:} Similar to a commit message, a
code change can be viewed as a sequence of words. This view, however,
overlooks the structure of code changes, as needed to distinguish between
changes to different files, changes in different hunks (contiguous
sequences of removed and added code), and different kinds of changes
(removals or additions). 
% Different from existing deep learning techniques
% working on source code~\cite{huo2017enhancing, lam2017bug}, PatchNet
% contains a novel hierarchical representation \jl{I'm not sure if this is
%   clear.  I'm not sure what is hierarchical?  The representation of the
%   data or the structure of the machine learning architecture?}
% to incorporate this structural information. 
To address this challenge, PatchNet contains a deep hierarchical structure that mirrors the hierarchical and sequential structure of patch code, making it distinctive from the existing deep learning models on source code~\cite{huo2017enhancing}.
For the code changes in a given patch, PatchNet outputs an embedding vector that represents the most salient features of these changes.

The code change module contains a {\em file module} that automatically
builds an embedding vector representing the code changes made to a given
file in the patch. Figure~\ref{fig:commit_code_model} shows the
architecture of the file module. This module takes as input two matrices
(denoted by ``--'' and ``+'' in Figure~\ref{fig:commit_code_model})
representing the removed code and added code for the affected file in a patch, respectively. These two matrices are passed to the \textit{removed code module} and the \textit{added code module}, respectively, to construct the corresponding embedding vectors by taking into account the structure of the removed or added code. 

The input of the \textit{removed code module} is a three dimensional matrix, indicating the removed code in the affected file of the given patch, denoted by $\mathcal{B}_r \in \mathbb{R}^{\mathcal{H}\times\mathcal{N}\times\mathcal{L}}$, where $\mathcal{H}$, $\mathcal{N}$, and $\mathcal{L}$ are the number of hunks, the number of removed code lines for each hunk, and the number of words of each removed code line in the affected file, respectively. This module constructs an embedding vector (denoted by $\textbf{e}_r$) representing the removed code in the affected file. The \textit{added code module} also takes as input a three dimensional matrix, indicating the added code in the affected file of the given patch. It follows the same architecture as the removed code module to construct an embedding vector (denoted by $\textbf{e}_a$) representing the added code in the affected file. 
These changes in the added and removed code are padded or truncated to have the same number of hunks ($\mathcal{H}$), number of lines for each hunk ($\mathcal{N}$), and number of words in each line ($\mathcal{L}$) for parallelization. Moreover, both modules also share the same vocabulary.

The two embedding vectors are then concatenated to represent the code changes in each affected file, i.e., $\textbf{e}_f = \textbf{e}_r \oplus \textbf{e}_a$. The embedding vectors of the code changes at the file level are then concatenated into a single vector representing all the code changes made by the patch.

\noindent \textbf{Classification module:} The classification module takes as input the commit message embedding vector ($\textbf{e}_m$) and the code change embedding vector ($\textbf{e}_c$) (see Figure~\ref{fig:patchnet}). These two vectors are then concatenated to form a single vector representing the patch, i.e., $\textbf{e} = \textbf{e}_m \oplus \textbf{e}_c$. The concatenated vector $\textbf{e}$ is passed to a fully-connected layer and an output layer, which computes a probability score for the patch. If an additional source of information is available, PatchNet can be easily extended by concatenating an additional information vector (denoted by $\textbf{e}_i$), collected from the data, to the commit message embedding vector and the code change embedding vector (i.e., $\textbf{e} = \textbf{e}_m \oplus \textbf{e}_c \oplus \textbf{e}_i$) to form a new vector representing the given patch.  

\noindent \textit{Parameter learning:} During the training process,
PatchNet uses adaptive moment estimation (Adam)~\cite{kingma2014adam} to minimize the regularized loss function~\cite{haykin2001kalman}. 
% Specifically \jl{Not sure what specifically means here}, 
PatchNet learns the following parameters: the word embedding matrices for commit messages and code changes, the filter matrices and bias of the convolutional layers, and the weights and bias of the fully connected layer and the output layer.  

%% file: preproc.tex
\section{Data selection and preprocessing}
\label{sec:data}

The user is responsible for selecting commits for training and annotating them according
to the chosen classification scheme.  For each commit, preprocessing is then applied to
the commit message and the code changes.  For the commit message, PatchNet
applies standard natural language preprocessing
techniques, such as stemming and stop word
elimination.  For the code changes, PatchNet detects the changed lines
using diff, and then expands these changes to include the complete
innermost enclosing simple statement, if any, or the header of a
conditional or loop, if the change occurs in such code.  Lines within the
changes are annotated as {\em error-checking code} (an {\tt if} test that
checks for failure of a previous operation), {\em error-handling code}
(code that performs cleanup in case of failure of a previous operation) or
{\em normal} (for everything else), reflecting one aspect of the semantics of the code.
% \jl{David suggests ``others'', but ``normal'' is what people
%   will see if they look at the file} (everything else).  
  PatchNet keeps the
names of called functions that are not defined in the current file and that occur
at least five times in the training dataset.  Other identifiers are
represented as a generic ``identifier'' token.  Because these preprocessing
steps require knowledge of the syntax of the language used by the source
code, PatchNet currently only supports C code.

%% file: usage.tex
\section{Usage}
\label{sec:usage}
In this section, we first describe the use of the preprocessor and then
illustrate several usage scenarios for the training and classification
process. Moreover, we describe a number of PatchNet's key hyperparameters
used during the training process.

\subsection{Preprocessing}

The preprocessing step is time consuming for large datasets, and is thus
separated from the rest of the training and classification process.  The main input to the
preprocessing step is a file containing a list of commit identifiers and their labels.
% (dummy labels can be used in the case of production \jl{not sure what is
%   the right word} data).  
The preprocessing step also needs to know the pathname of the git tree
containing the commits and a prefix used to construct the names of the various output
files.  The output is a pair of files, containing the patch data and a
dictionary for interpreting the patch data.  Only the former needs to be
provided to the subsequent training and classification process.  A typical
command line is:

{\small\begin{verbatim}
getinfo --commit_list commit_list_file
  --git /path/to/git -o training_data
\end{verbatim}}

\subsection{Scenario I - Simple Command}
\label{sec:simple_command}
PatchNet is implemented in Python 2.7 with Tensorflow
1.4.1,\footnote{\url{https://www.tensorflow.org/}} scikit-learn
0.19.1,\footnote{\url{http://scikit-learn.org/stable/}} and
numpy{1.14.3}.\footnote{\url{http://www.numpy.org/}} 
% We chose Python due to
% its simplicity and convenience when deploying deep learning libraries
% (i.e., Tensorflow). 
PatchNet performs deep patch classification in two phases: the training phase and the prediction phase. 

In the training phase, PatchNet takes the command-line arguments \texttt{--train} indicating the training phase, \texttt{--data} indicating the path of 
a list of labeled patches, and \texttt{--model} indicating the name of the output folder in which to put the model. A sample command that trains a model is as follows:

{\small\begin{verbatim}
python PatchNet.py --train
  --data training_data.out --model patchnet
\end{verbatim}}

% \jh{David's comment: This example is concrete (i.e.,
%   ``training\_data.out''), while the previous example ``getinfo'' is
%   abstract (i.e., ``$<$commit list file$>$''). Could we standardize the way
%   we present the commands? \textbf{Julia: can you please address his
%     comment?}}

\noindent
This command instructs PatchNet to train a model using the data \texttt{training\_data.out} with the default
hyperparameters. In the default setting, the dimension of the embedding
vectors, the number of filters, and the number of hidden layers are set
to 32, 32, and 10, respectively. The dropout for the training process, the
regularization error, and the learning rate are set to 0.5, 1$e-$5, and 1$e-$4,
respectively. PatchNet automatically creates a new folder with the name
\texttt{patchnet} (or empties a folder with that name, if it exists)
and saves a learned model, consisting of three files, in the folder. For parallelization, the number of changed files, the number of hunks for each file, the number of lines for each hunk, the number of words of each removed or added code are set to 5, 8, 10, and 120, respectively. 

% \jh{David's comment: What is the name of the model file? How if multiple model files are in the folder? What happen? Would it crash?}
% \jg{In our example, if the name's model is ``patchnet'', Tensorflow library will automatically create a folder name ``patchnet''. Inside the folder ``patchnet'' is three files: ``patchnet.data'', ``patchnet.index'', ``patchnet.meta''. The three files represent the model named ``patchnet''. If users are familiar with Tensorflow library, I think they should know it.}

In the prediction phase, PatchNet takes the command-line arguments
\texttt{--predict} indicating the prediction phase, \texttt{--data} indicating the path of 
a list of unlabeled patches, and \texttt{--model} indicating the name of
the folder containing the model. The following is a sample command to collect a list
of prediction scores for a set of unlabeled patches: 

{\small\begin{verbatim}
python PatchNet.py --predict
  --data test_data.out --model patchnet
\end{verbatim}}

PatchNet has been developed and tested on the Linux platform. However, we believe that PatchNet can be employed on other platforms such as Windows or MacOS if all the necessary libraries are installed (i.e., Tensorflow, scikit-learn, and numpy). More details about PatchNet's installation instructions can be found at \url{https://github.com/hvdthong/PatchNetTool}.

\subsection{Scenario II - Tuning PatchNet's Hyperparameters}
\label{sec:tunning_parameters}

\begin{table}[t!]
  \centering
  \caption{The key hyperparameters of PatchNet}
  \label{tab:params}%
    \begin{tabular}{|l|l|}
    \hline
    \textbf{Hyperparameters} & \textbf{Description} \\
    \hline
    \hline
    \texttt{--data\_type} & Type of data (commit messages, code  \\ 
    & change, or both) used to construct a model. \\
    & Default: both. \\
    \hline
    \texttt{--embedding\_dim} & Dimension of embedding vectors. \\
    & Default: 32. \\
    \hline
    \texttt{--filter\_sizes} & 
            Sizes of filters used by the convolutional\\
    & layers. Default: ``1, 2''.  \\
    \hline
    \texttt{--num\_filters} & Number of filters. Default: 32. \\
    \hline
    \texttt{--hidden\_layers} & Number of hidden layers. Default: 16.  \\
    \hline
    \texttt{--dropout\_keep\_prob} & Dropout for training PatchNet. Default: 0.5. \\
    \hline
    \texttt{--l2\_reg\_lambda} & Regularization rate. Default: 1$e-$5. \\
    \hline
    \texttt{--learning\_rate} & Learning rate. Default: 1$e-$4. \\
    \hline
    \texttt{--batch\_size} & Batch size. Default: 64. \\
    \hline
    \texttt{--num\_epochs} & Number of epochs. Default: 25. \\
    \hline
    \end{tabular}%
\end{table}%

Table~\ref{tab:params} highlights the key hyperparameters that control how
PatchNet trains its model.  The \texttt{--data\_type} flag allows the user
to choose what information to include from the patches (i.e., commit messages,
code changes, or both). The user can also change other hyperparameters of
the model such as the dimension of embedding vectors (i.e.,
``\texttt{--embedding\_dim}''), number of filter sizes (i.e.,
``\texttt{--filter\_sizes}''), number of filters
(i.e.,``\texttt{--num\_filters}''), etc. The following example illustrates
how to execute PatchNet with custom settings:

{\small\begin{verbatim}
python PatchNet.py --train
  --data data.out --model patchnet
  --embedding_dim 128 --filter_sizes "1,2"
  --num_filters 64
\end{verbatim}}

% \noindent \texttt{python PatchNet.py --train}

% $\quad$ \texttt{--data ./data.out} 

% $\quad$ \texttt{--model patchnet} 

% $\quad$ \texttt{--embedding\_dim 128} 

% $\quad$ \texttt{--filter\_sizes ``1,2''}

% $\quad$ \texttt{--num\_filters 32}

\section{Stable Patch Identification}
\label{sec:exp}

We have applied PatchNet to the problem of identifying Linux kernel
bug-fixing patches that should be backported to previous stable versions.
Based on a set of 42,408 stable patches and 39,995 non-stable
patches drawn from Linux kernel versions from v3.0 (July 2011) to v4.12
(July 2017), we trained and tested the PatchNet model using its default hyperparameters following 5-fold cross-validation. Considered patches are limited to 100 lines of changed code, following the Linux kernel stable patch guidelines.  Non-stable patches in the dataset are chosen to have the same size properties (number of files and number of changes lines) as the stable ones. 
 
Table~\ref{tab:components} shows the performance (i.e., accuracy, precision, recall, and F1) on this problem of LPU+SVM~\cite{tian2012identifying} and three variants of PatchNet: PatchNet-C, PatchNet-M, and PatchNet. PatchNet-C uses only code change information while PatchNet-M uses only commit message
information. PatchNet uses both commit message and code change information.
Accuracy, precision, recall, F1, and AUC for the stable patch
identification problem drop by 15-20\% if we ignore code changes and
14-17\% if we ignore the commit message, showing the interest of a model that
incorporates both kinds of information. Table~\ref{tab:components} also shows that PatchNet outperforms the state-of-the-art baseline (i.e., LPU+SVM) for the task of stable patch identification. 

\begin{table}[t!]
  \centering
  \caption{Contribution of commit messages and code changes to PatchNet's performance}
    \begin{tabular}{|l|c|c|c|c|c|}
    \hline
          & \textbf{Accuracy} & \textbf{Precision} & \textbf{Recall} & \textbf{F1}    & \textbf{AUC} \\
    \hline
    \hline
    LPU+SVM & 0.731 & 0.751 & 0.716 & 0.733 & 0.731 \\
    \hline
    PatchNet-C & 0.722 & 0.727 & 0.748 & 0.736 & 0.741 \\
    \hline
    PatchNet-M & 0.737 & 0.732 & 0.778 & 0.759 & 0.753 \\    
    \hline
    PatchNet & \textbf{0.862} & \textbf{0.839} & \textbf{0.907} & \textbf{0.871} & \textbf{0.860} \\
    \hline
    \end{tabular}%
  \label{tab:components}\vspace{-0.1cm}
\end{table}%

%% file: conclusion.tex
\section{Conclusion}
\label{sec:conclusion}

In this work, we present PatchNet, a tool that learns a semantic representation of patches for classification purposes. PatchNet contains a deep hierarchical structure that mirrors the hierarchical and sequential structure of commit code, making it distinctive from the existing deep learning models on source code. We have demonstrated the tool's applicability in identifying stable patches in Linux kernel. We encourage future researchers to benefit from PatchNet by applying it to other tasks that can be mapped to a patch classification problem. PatchNet is open-source and can be run from the command line with simple options. 
% PatchNet's implementation and our dataset used to evaluate PatchNet's applicability for stable patch identification task are publicly available at:~\url{https://github.com/hvdthong/PatchNetTool}.

%% file: ack.tex
\vspace{0.2cm}\noindent{\bf Acknowledgement.} This research was supported by the Singapore National Research Foundation (award number: NRF2016-NRF-ANR003) and the ANR ITrans project.